\documentclass[epj]{webofc}
\usepackage[varg]{txfonts}   
\begin{document}
%

\title{e-$\mu$ discrimination at high energy in the JUNO detector}
%
%

\author{\textit{Giulio} Settanta\inst{1,2}\fnsep\thanks{\email{giulio.settanta@uniroma3.it}},
        \textit{Stefano Maria} Mari\inst{1,2},
        \textit{Cristina} Martellini\inst{1,2} \and
        \textit{Paolo} Montini\inst{1,2}
        , on behalf of the JUNO Collaboration
}

\institute{Dipartimento di Matematica e Fisica, Università degli Studi Roma Tre, Via della Vasca Navale 84, 00146 Rome, Italy
\and
           INFN Sezione di Roma Tre, Via della Vasca Navale 84, 00146 Rome, Italy
          }

\abstract{%
  Cosmic Ray and neutrino oscillation physics can be studied by using atmospheric neutrinos. JUNO (Jiangmen Underground Neutrino Observatory) is a large liquid scintillator detector with low energy detection threshold and excellent energy resolution. The detector performances allow the atmospheric neutrino oscillation measurements. In this work, a discrimination algorithm for different reaction channels of neutrino-nucleon interactions in the JUNO liquid scintillator, in the GeV/sub-GeV energy region, is presented. The atmospheric neutrino flux is taken as reference, considering $\overset{(-)}{\nu_\mu}$ and $\overset{(-)}{\nu_e}$. The different temporal behaviour of the classes of events have been exploited to build a time profile-based discrimination algorithm. The results show a good selection power for $\overset{(-)}{\nu_e}$ CC events, while the $\overset{(-)}{\nu_\mu}$ CC component suffers of an important contamination from NC events at low energy, which is under study. Preliminary results are presented.
}
\maketitle
\section{Introduction}
\label{intro}
The origin and properties of Cosmic Rays (CRs) are still matter of study by a number of experiments. A fraction of the energy of an air shower, resulting after a CR interaction in the atmosphere, is carried by neutrinos. The atmospheric $\nu$ flux is almost entirely composed of $\overset{(-)}{\nu_\mu}$ and $\overset{(-)}{\nu_e}$ and spans many decades in energy, from the MeV up to the PeV scale. Neutrinos travelling across the Earth can be also used as a probe to study flavor oscillation effects in matter. After the experimental results in past years \cite{atmoFrejus, atmoSK, atmoIC3, atmoIC4}, further contributions to the atmospheric $\nu$ spectrum will come from the next generation of neutrino detectors, operating in the next decade. JUNO is a large liquid scintillator detector with low energy threshold and excellent energy resolution ($\sim 3\%/\sqrt{E~[MeV]}$), under construction in China \cite{YBJuno}. Its core consists of a $\sim$36 m diameter acrylic sphere, which will be filled with 20kt of liquid scintillator. The light produced in $\nu$ interactions will be collected by a double-system of photosensors: 17.000 20" PMTs and 25.000 3" PMTs. Given the large fiducial volume and the excellent energy resolution, JUNO will be able to detect atmospheric neutrinos.\\
\noindent A crucial part in the atmospheric $\nu$ study is the correct identification of the original $\nu$ flavor. Above $\sim$100 MeV neutrinos can be assumed to interact with nucleons only. This implies that hadronic particles are always produced in the final state. In the case of a charged-current (CC) interaction, the corresponding charged lepton is also produced (either $\nu_\mu N \rightarrow \mu X$ or $\nu_e N \rightarrow e X$). In the case of a neutral-current (NC) interaction, only secondary hadrons and the scattered neutrino are in the final state.
\section{The Monte Carlo simulation}
\label{sec-MC}
\begin{figure}[ht]
\centering
\includegraphics[width=\textwidth/2]{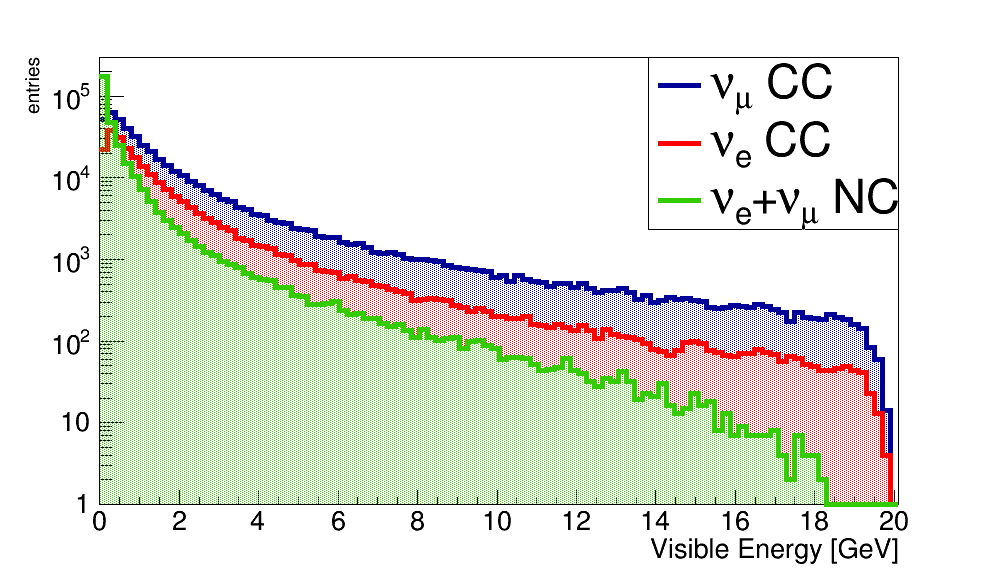}
\includegraphics[width=\textwidth/3]{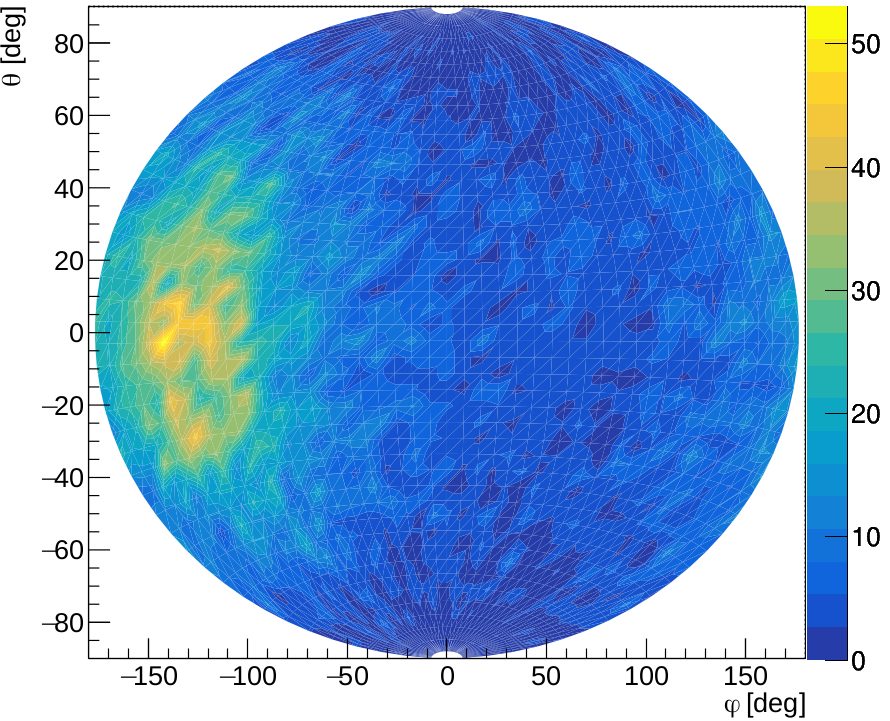}
\caption{Left: visible energy of the MC sample, for $\nu_\mu$ CC (blue), $\nu_e$ CC (red) and NC events (green); right: map of the hits on the 3'' PMT system after a E = 378 MeV $\nu_\mu$ CC interaction at positions X = 4.9 m, Y = 7.2 m, Z = -0.05 m, assuming the detector centre at (0,0,0).}
\label{MC-1}       
\end{figure}

Since the JUNO detector is still under construction, all the analysis so far relies on Monte Carlo (MC) simulations only. The MC production proceeds in several steps:
\begin{itemize}
    \item Generation of the atmospheric $\nu$ flux at the JUNO location, according to the predictions from \cite{hkkm15}. The model provides also the zenith- and azimuth-angle flux dependance.
    \item Interaction of neutrinos with the JUNO liquid scintillator, by means of the \emph{GENIE Neutrino Monte Carlo Generator} \cite{GENIE}. The events have been considered up to 20 GeV and consist of $10^6~ \overset{(-)}{\nu_\mu} + \overset{(-)}{\nu_e}$ events (fig. \ref{MC-1} - left). The output of the simulation is the full list of secondaries and their associated properties (ID, momentum, direction ...).
    \item Propagation of secondaries in the scintillator by means of a \emph{GEANT4-}based simulation \cite{Sniper}, which includes energy losses, photon production and propagation, photon-PMT cathode interaction and photo-electrons (PE) generation. This last step includes 50k events only. Since JUNO includes an active volume of 20000 tons and more than 40000 PMTs, running a full simulation is both CPU and storage consuming, which is a limitation in the number of simulated events.
\end{itemize}
\noindent At the end of the simulation, a map of the hits on the detector is available, distributed in time and position (fig. \ref{MC-1} - right). 
\section{Flavor identification}
\begin{figure}[ht]
\centering
\includegraphics[width=\textwidth]{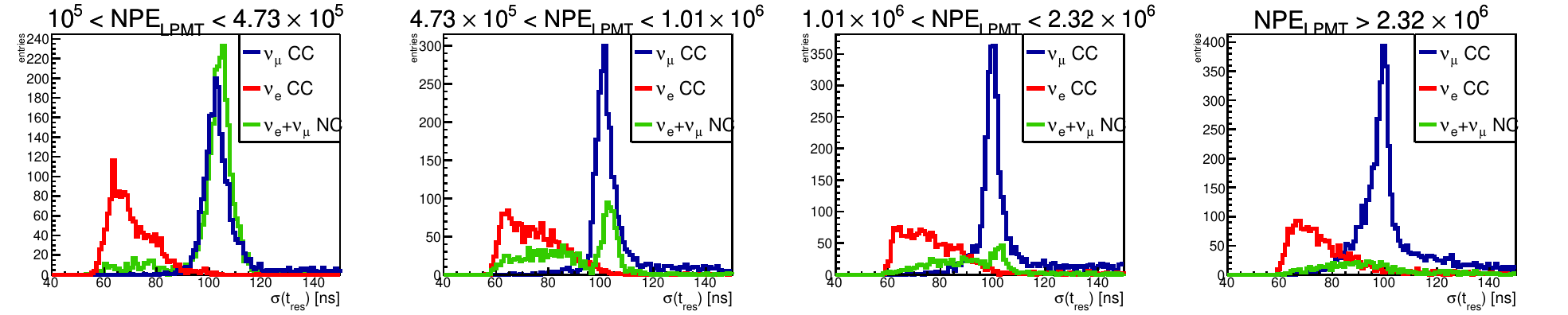}
\caption{Distribution of $\sigma(t_{res})$ for $\nu_\mu$ CC (blue), $\nu_e$ CC (red) and NC events (green). The NPE range is reported above the figures.}
\label{tres}
\end{figure}

\begin{figure}[ht]
\centering
\includegraphics[width=0.49\textwidth]{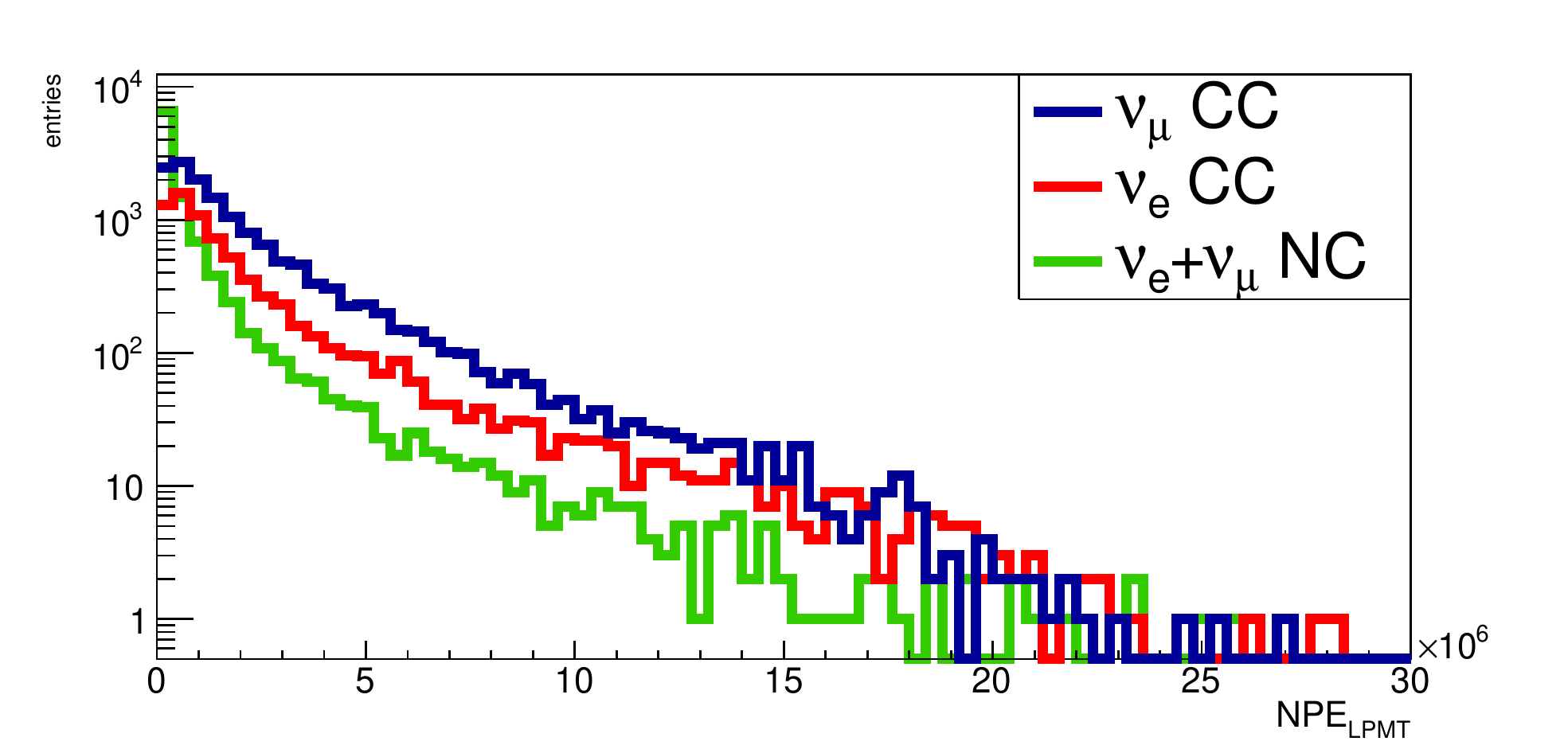}
\includegraphics[width=0.49\textwidth]{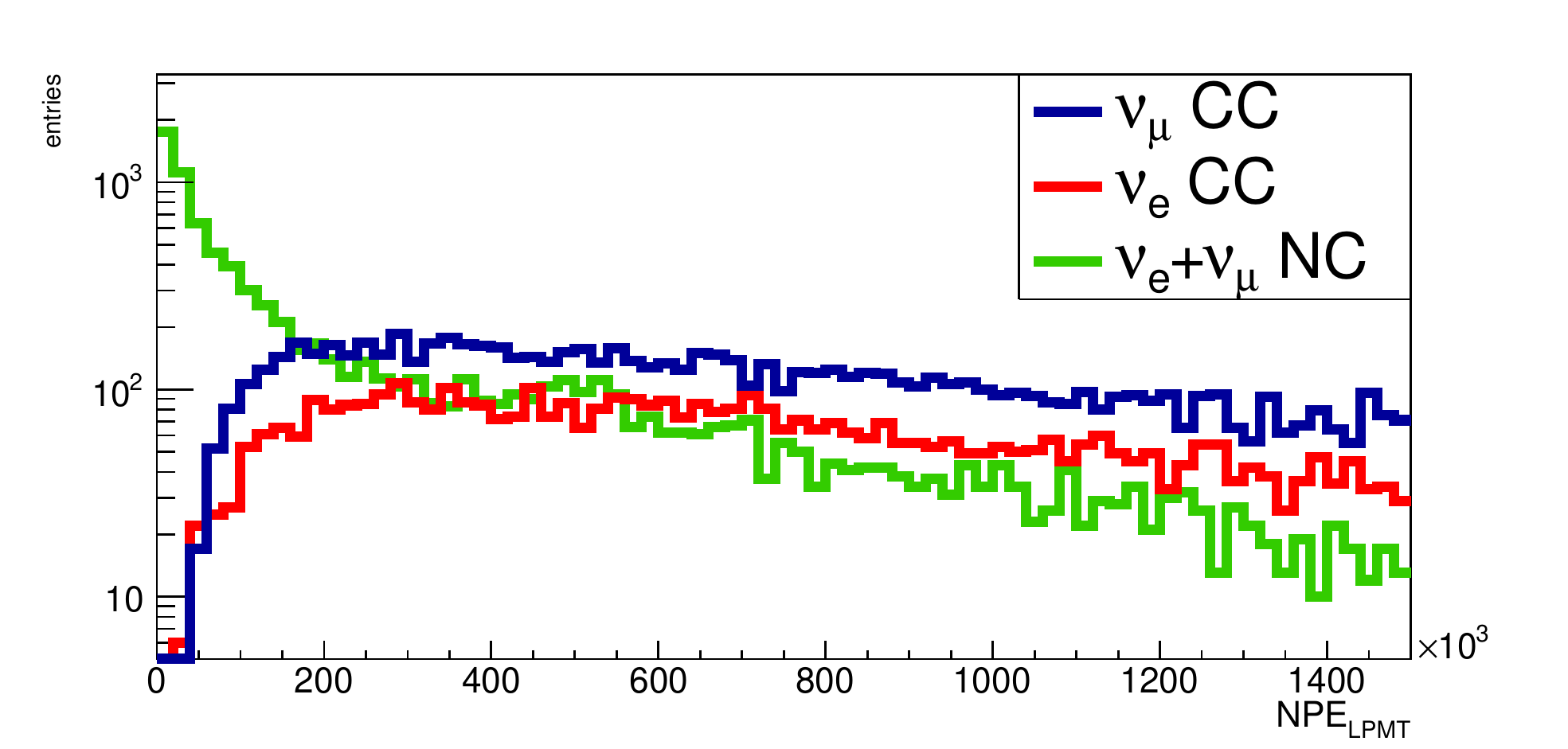}
\caption{Left: distribution of the NPE collected by the 20" PMT system, according to the the interaction channel (same legend of fig. \ref{tres}); right: same distribution, in the lower energy band.}
\label{fig:npe}
\end{figure}

Since in CC interactions there is always an electron or a muon in the final state, they constitute the preferred channel for the neutrino flavor identification. Muons with energy > 1 GeV travel in general for a longer distance inside the detector with respect to electrons. Additionally, muons can decay inside the scintillator volume, giving a delayed energy release at low energy too. The above differences make $\nu_\mu$ CC events more elongated in time with respect of $\nu_e$ CC events, which indeed appear more point-like. Hadronic particles, which constitute NC events, have in general a long-living energy release, because of interactions and decays. Furthermore, since hadrons add in general late energy releases to all events, they also make more similar $\nu_\mu$ and $\nu_e$ CC event topologies.\\
The different temporal behaviour of the classes of events has been exploited to build a time profile-based discrimination algorithm. The 3'' PMT system has been used in the algorithm, since it is more accurate in time measurement. To reproduce realistic Time-Transit Spread (TTS) effects, an artificial smearing has been applied to the true hit time over every PMT. A gaussian function with $\sigma = 4~ns$ has been used. A further selection on the smeared vertex is applied, in order to avoid edge effects, requiring it to be be within 16 m from the detector centre. The hit time-residual is defined for each PMT as:
\begin{equation}\label{eq:tres}
t_{res}^i = t_{hit}^i - \left( \frac{c}{n \cdot R_V^i} \right)
\end{equation}
where $c/n$ is the speed of light inside the scintillator and $R_V^i$ is the distance between the i-th PMT and the interaction vertex. In order to reproduce conservatively the uncertainty on the vertex reconstruction at the energy considered, a smearing with a $\sigma = 1~m$ gaussian function has been applied to all events. Since $\nu_\mu$ and $\nu_e$ CC events result in different light production duration, the RMS of the $t_{res}$ is used as a discrimination variable (here called $\sigma(t_{res})$). In fig. \ref{tres} the $\sigma(t_{res})$ distribution is reported for the three populations: $\nu_\mu$ CC, $\nu_e$ CC and NC events. The variable is also reported for 4 different bins of number of PEs (NPE) collected by all 20'' PMTs, selected in order to have equal statistics in each bin.
From fig. \ref{tres}, a good separation of the $\nu_e$ CC component is evident, in all the energy bins. On the contrary, the $\nu_\mu$ CC and the NC components are almost overlapped on the whole energy range. The NC component contribution, however, becomes less significant at high energy. The spectral distribution of the MC sample, according to the interaction channel, is shown in fig. \ref{fig:npe} and reflects the sub-dominance of the NC component at high energy.

\begin{figure}[ht]
\centering
\includegraphics[width=0.49\textwidth]{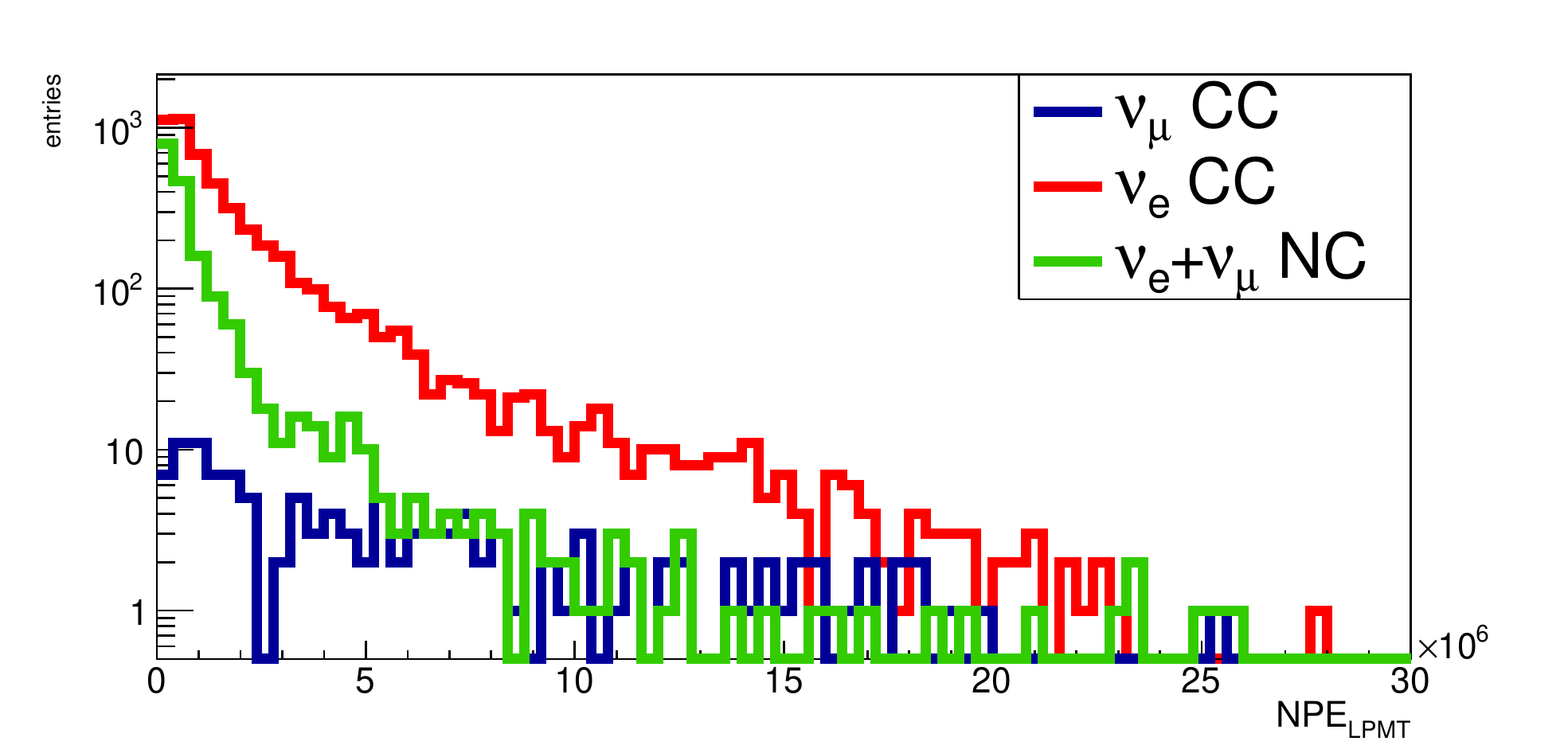}
\includegraphics[width=0.49\textwidth]{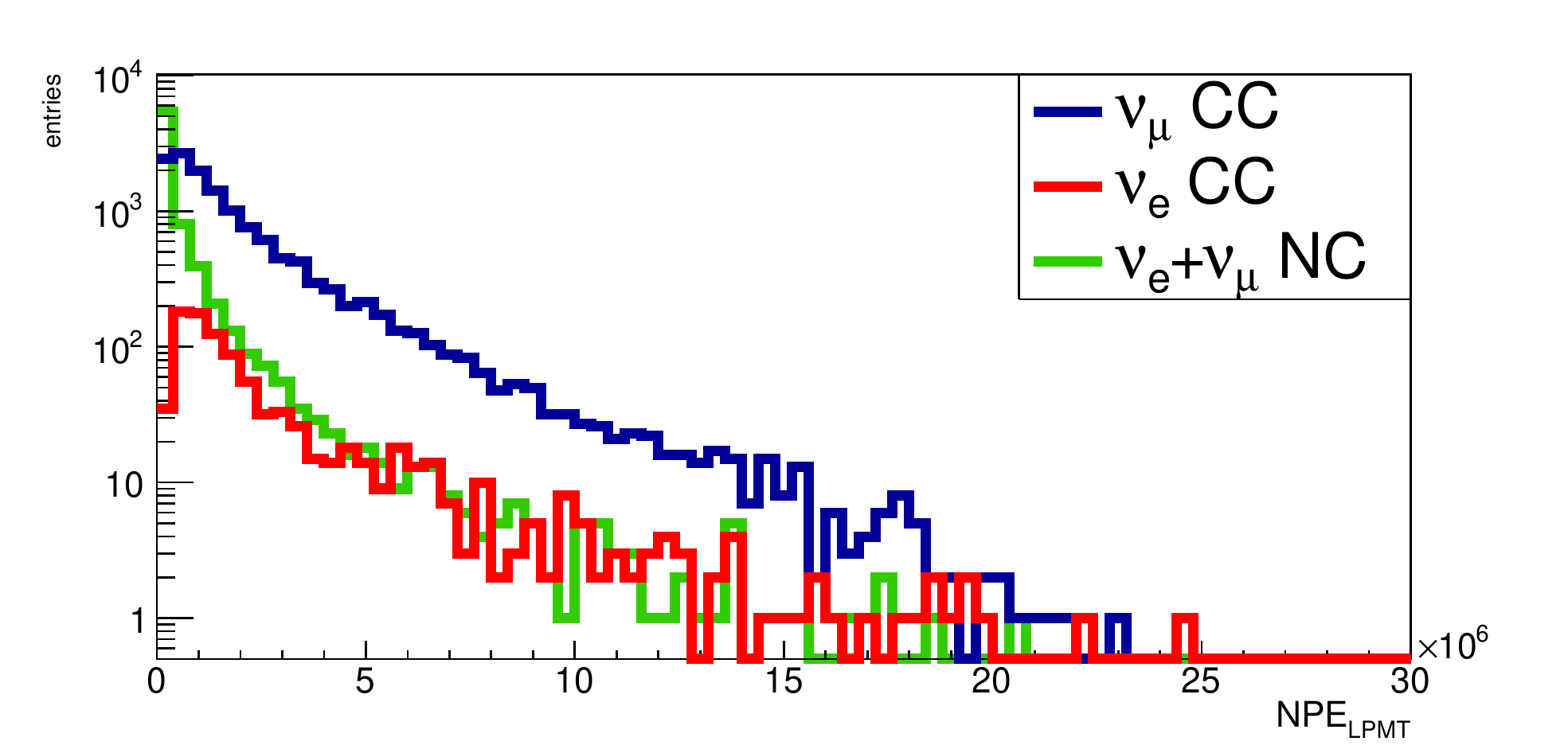}
\caption{Left: distribution of the NPE collected by the 20" PMT system, of events with $\sigma(t_{res}) < 80~ns$, according to the the interaction channel (same legend of fig. \ref{tres}); right: same distribution, for $\sigma(t_{res}) > 90~ns$.}
\label{cuts}
\end{figure}

\noindent Given the distributions reported in fig. \ref{tres}, a sharp cut on the $\sigma(t_{res})$ is applied to select the populations of interest: $\sigma(t_{res}) < 80~ns$ for $\nu_e$ CC and $\sigma(t_{res}) > 90~ns$ for $\nu_\mu$ CC. In fig. \ref{cuts} the NPE distribution is reported, after the above cuts.
Preliminary studies on the effects of the cuts have been evaluated on an independent MC sample of 10k events. Considering a range between $10^5$ and $10^7$ NPE, the $\sigma(t_{res}) < 80~ns$ cut keeps 70\% of the $\nu_e$ CC events with a contamination around 17\%. In the range above $10^6$ NPE, the $\sigma(t_{res}) > 90~ns$ cut keeps 93\% of the $\nu_\mu$ CC events with a contamination around 17\%. These performances represent a preliminary result in the separation of the samples and are under study for further optimization.
\section*{Conclusions}
An event time-profile algorithm is used for the e/$\mu$ separation after atmospheric neutrinos interaction in the JUNO detector. Preliminary results are reported on the poster and show a good selection power for $\nu_e$ CC events. The $\nu_\mu$ CC is still contaminated by an important fraction of NC events at low energy and is under study. By evaluating the cuts on an independent MC sample, in a range between $10^5$ and $10^7$ NPE, the $\sigma(t_{res}) < 80~ns$ cut keeps 70\% of the $\nu_e$ CC events with a contamination around 17\%. In a range above $10^6$ NPE, the $\sigma(t_{res}) > 90~ns$ cut keeps 93\% of the $\nu_\mu$ CC events with a contamination around 17\%. The above results are still preliminary and are under optimization.

%

\end{document}